\documentclass[12pt]{revtex4} \usepackage[dvips]{graphicx,color}
\def\bea{\begin{eqnarray}} \def\eea{\end{eqnarray}}
\def\be{\begin{equation}}
\def\ee{\end{equation}}
\def\nn{\nonumber}

\def\g{\gamma}
\def\G{\Gamma}

\def\m{\mu}
\def\n{\nu}

\def\t{\tau}

\def\e{\epsilon}

\begin{document}

\title{Quantum gravity effects on unstable orbits in Schwarzschild space-time}
\vspace{1cm}

\author{\sc Arundhati Dasgupta}
\affiliation{
  Department of Physics and Astronomy, University of Lethbridge, 4401 University Drive, Lethbridge, T1K 3M4}
\email{arundhati.dasgupta@uleth.ca}
\begin{abstract} 
We study semiclassical corrections to the Schwarzchild metric and their effects on unstable orbits.
\end{abstract}
\maketitle

\section{Introduction}
It was observed in \cite{cornish} that certain orbits for particles in a Schwarzschild space-time were `unstable' to perturbations. A slight change in the
initial conditions or perturbations of these orbits have exponential growth, in other words, one can find Lyapunov exponents. 
We study these orbits and the modifications of the Lyapunov exponents under effects of quantum gravity. 


We find semiclassical corrections to the metric using coherent states in Loop Quantum Gravity (LQG), described in \cite{thiem1,thiem2,adg1,adg2}. The corrected metric
is then used to find unstable orbits. The Lyapunov exponents for circular orbits in the corrected metric are determined using the method of \cite{cornish}. It is found that non-zero Lyapunov exponents occur even for
normally classically stable orbits, but these cause insignificant effects for macroscopic astrophysical black holes. The time scale for the
instabilities becomes significant ($\sim 1$ second (s)) for primordial black holes with horizon radius of the order of $10^{14}$ Planck length ($l_p$). It is then
shown that this instability will not affect `gravity waves' from these primordial black holes. However, this `new' 
semiclassical instability might affect black hole mergers and other phenomena like quasinormal ring down from these
primordial black holes \cite{quasi,merge}. Next, we find the growth of the semiclassical corrections to classical circular orbit radii
as per the classical Lyapunov exponents. This effect is finite
for astrophysical black holes. The deviations grow to the order of 1 cm in a time of 1 second for black holes with horizon radius $10^{42} l_p$ which thus includes astrophysical black holes.
However, in an actual accreting disc it will be difficult to isolate these
instabilities from external matter perturbations.

In the next section semiclassical corrections to the metric using coherent states of LQG are described. Section III finds the effects
 of these corrections on the unstable orbits. The last section describes some physical systems where one can detect this effect of quantum gravity. 
\section{Semiclassical corrections}
We discuss semiclassical states which are special as expectation values
of operators are closest to their classical values in these states. In case of quantum gravity such special states were identified in LQG framework in \cite{thiem1,thiem2} using SU(2)
coherent states defined by Hall. 

LQG is a formulation rooted in canonical gravity, where one takes a suitable time slicing of a space-time by identifying a fiducial time like direction. The induced metric in the
time slices is denoted by $q_{ab}$ (a,b=1..3) and the extrinsic curvature with which this slice is embedded in four dimensions is denoted by $K_{ab}$. 
The theory of loop quantum gravity is defined in terms of the triads or the square root of the metric $q_{ab} = e_a^I e_b^I$ and the `spin connection' $\G^I_a = \e^{IJK}e^b_J\nabla_a e_{bK}$. 
$I=1,2,3$ represents an internal SU(2) index. A redefinition of the variables in terms of tangent space densitised triads $E_a^I$
and a corresponding gauge connection $A_a^I$ simplifies the quantisation considerably. 
\be
A_a^I =\G_a^I - \beta K_{ab} e^I_b \ \ \ \ E^a_I = \frac1{\beta} ({\rm det} \ e) e^a_I
\label{defn} 
\ee
($e_a^I$ are the usual triads, $\G_a^I$ the associated spin connection, and $\beta$ the one parameter ambiguity which remains named as the Immirzi
parameter and is set to 1 for the discussions of the paper.)

The `holonomy' $h_e(A)= {\cal P} \exp(\int_e A.dx)$ and the
momentum $P_{e}^I= \int_{S_{e}}*E^I_a$ as integrals over an edge $e$  and 2-surface $S_{e}$ (which is intersected by the edge $e$ at one point) are defined as `discretisations' of the gauge connection and the triads. The discrete variables give a well defined Poisson bracket. 
One can find a kinematic Hilbert space which carries a representation of this algebra. A semiclassical state $\psi^{\tilde t}$ defined in the kinematic Hilbert space is
given as a function of $h_e$, $g_e= e^{i T^I P^I_e} h_e$  ($T^I$ are the three $2\times 2$ SU(2) generator matrices) and ${\tilde t}$ 
(a semiclassical parameter). For the purposes of this discussion the semiclassical parameter is taken as ${\tilde t}=\frac{l_p^2}{r_g^2}= 10^{-2n} $, where $r_g=10^n l_p$ is the radius of the horizon of the Schwarzschild black hole. $\tilde t\rightarrow 0$ represents the `classical limit' and the wavefunction
is given by 
\be
\psi^{\tilde t} = \sum_j \ (2j+1) e^{-\tilde t j(j+1)/2} \chi_j(g_e h_e^{-1})
\ee
(j=0,$\frac12$ ,1.., and $\chi_j$ is the character of the jth SU(2) representation of the matrix $g_e h_e^{-1}$). The wavefunction is nicely peaked at the classical values
of $h_e$ and $P_{e}^I$. In the particular coordinates used to describe the embedding of the edges in the classical metric, they are labelled by $e_a$, where $a$ labels the three space dimensions. The fluctuations over the classical values can be obtained as a series in powers of ${\tilde t}$. ( We use the $\tilde t$ notation to avoid confusing this with the time coordinate.) Thus
\be
<\hat P_{e_a}^I>= \frac{<\psi^{\tilde t}|\hat P_{e_a}^I |\psi^{\tilde t}>}{||\psi^{\tilde t}||}= P_{e_a}^I\left[ 1 + {\tilde t} \tilde f(P)\right]
\ee
where $\tilde f(P)$ is a function and measures the first order correction to the classical value. The details of the function $\tilde f(P)$ have been calculated previously in \cite{thiem2,adg1}.
Here we redo the calculations to ascertain the exact nature of the corrections. The action of the operator $\hat P^I_{e_a}$ on the coherent state is given by 
\be
\hat P^I_{e_a}\psi^{\tilde t}= \frac{\iota \tilde t}{2} \left(\frac{d}{d\gamma}\right)_{\g=0} \psi^{\tilde t}\left(e^{\g T^I} h_e\right)
\ee
($\gamma$ being a real parameter).
The numerator of the expectation value of the momentum operator $\hat P_{e_a}^I$   
\be
<\psi^{\tilde t} |\hat P_{e_a}^I |\psi^{\tilde t}>= \frac{\iota \tilde t}{2} \left(\frac{d}{d\g}\right)_{\g=0} \sum_j (2j+1) e^{-{\tilde t} j(j+1)} \frac{\sinh (2j+1) z}{\sinh z}
\ee
where $\cosh (z)= \frac12 {\rm Tr} (e^{-\g T^I} g \bar g^T)$, $j=0,1/2,1$ labels the SU(2) angular momentum.
By defining $n=2j+1$, one can rewrite the sum as
\be
=\frac{\iota \tilde {t}}{2}\left(\frac{d}{d\g}\right)_{\g=0} \frac{e^{\tilde {t}/4}}{\sinh (z)}\sum_{n=0}^{\infty} \ n \ e^{-\tilde {t} n^2/4} \sinh (n z)
\ee
In the limit $t\rightarrow 0$ the above sum is converted into an Riemannian integral
\bea
&=& 2 \iota \left(\frac{d}{d\g}\right)_{\g=0} \frac{e^{\tilde{t}/4}}{\sinh(z)} \int_{0}^{\infty} x e^{-x^2}\sinh(\bar z x) dx \\ 
&=& 2 \sqrt{\pi}\iota \left(\frac{d}{d\g}\right)_{\g=0} \frac{e^{t/4}}{4\sinh(z)} \bar z \exp\left(\frac{\bar z^2}4\right)  
\eea
where $\bar z= 2 z/\sqrt{\tilde t}$. Using the approximations of \cite{thiem2}, one writes $z= P_{e_a} + \delta$ where $P_{e_a}= \sqrt{P_{e_a}^I P_{e_a}^I}$ and $\delta = -\g{\rm Tr}(T^I g \bar g^T)/(2 \sinh(P_{e_a}))$ to first order in $\gamma$, and thus
one can easily take the derivative wrt $\gamma$. The norm of the wave function is $||\psi^{\tilde t}||= 2\frac{P_{e_a}}{\sinh(P_{e_a})}\sqrt{\pi} e^{t/4} e^{P_{e_a}^2/t}$, 
and 
\be
P^I_{e_a}=\frac{-\iota}{2}\frac{{\rm Tr}(T^I g\bar g^T) P_{e_a}}{\sinh P_{e_a}}
\ee
Thus after some algebra
\be
<P^I_{e_a}> = P^I_{e_a}\left[ 1 + \frac{\tilde t}{P_{e_a}} \left(\frac1{P_{e_a}} - \coth (P_{e_a})\right)\right]
\ee
Interestingly in this approximation, the correction is proportional to the original value of $P^I_{e_a}$. The corrections are negative or positive as per the sign of $P^I_{e_a}$. Thus
as per the above \be \tilde f(P)= \frac{1}{P_{e_a}}\left(\frac1{P_{e_a}} - \coth (P_{e_a})\right)\ee
The sign of which is negative for all values of $P_{e_a}$

In \cite{adg1}, I used a coherent state defined on a flat slicing of the Schwarzschild metric, and one starts with a metric defined in Lemaitre coordinates:
\be
ds^2 = -d\tau^2 + \frac{dR^2}{\left[\frac{3}{2r_g}(R \pm \t)\right]^{2/3}} + \left[\frac{3}{2}(R \pm \t)\right]^{4/3} r_g^{2/3}(d\theta^2 +\sin^2\theta d\phi^2)
\label{lema}
\ee
The $+$ sign in the above metric corresponds to a white hole with signals emitted from the centre $r=0$, but $-$ sign corresponds to a black hole with material falling to $r=0$.
Of course in a collapsing situation, the $-$ sign should be solely relevant.
The $R,\tau$ coordinates are related to the Schwarzschild coordinates $t,r$ using the following transformations:
\be
\sqrt{\frac{r}{r_g}} dr  = (dR \pm d\tau) \ \ \ 
dt = \frac{1}{1-f'}\left(d\tau \pm f' dR \right)\ \ \ f'  = \frac{r_g}{r}
\label{tran1}
\ee
The constant $\t=\t_c$ surface then has a flat metric. For the either of the signs used in the Lemaitre metric, the induced metric in the $\tau=$ constant slice
is $ds^2= dr^2 + r^2 (d\theta^2 + \sin^2\theta d\phi^2)$ where $r^{1/2}r_g^{-1/2} dr=dR$ is the transformed coordinate on the slice. The extrinsic curvature of the Lemaitre coordinate can also be similarly transformed to the induced 
coordinate. In the set of induced coordinates, the momenta $P_{e_r}^I$ (momentum in the radial direction), $P_{e_\theta}^{I}$ (momentum in the $\theta$ direction) and $P_{e_\phi}^I$ 
(momentum in the $\phi$ direction) were calculated in \cite{adg1}. The two surface bits used to compute the momenta were bits of 2-spheres, and in the limit the area
of these bits went to zero, $P_{e_a}^I= S_{e_a} E_{e_a}^I$, where $S_{e_a}$ is the area of the two surface.
Given that, $q q^{ab} =E^{aI} E^{bI} = \frac{P^I_{e_a}}{S_{e_a}}\frac{P_{e_b}^I}{S_{e_b}}$ we get $q= {\rm det} \frac {P^I_{e_a}}{S_{e_a}}=P$.
Thus
\be
q^{ab}= \frac{1}{P} \frac{P^I_{e_a}}{S_{e_a}}\frac{P_{e_b}^I}{S_{e_b}}
\ee
Calculating 
\be
<\psi^{\tilde t}|\frac{\hat P^I_{e_a}}{S_{e_a}}\frac{\hat P_{e_b}^I}{S_{e_b}}|\psi^{\tilde t}>
\ee
should be enough to calculate corrections to the metric.

For the specific purpose of calculating corrections to the unstable orbits, we find the corrections to the radial metric.
\be
q^{rr}= \frac{1}{P}\left(\frac{P_{e_r}}{S_{e_r}}\right)^2 \left(1 + 2 \tilde t \tilde f\left(\frac{P_{e_r}}{S_{e_r}}\right)\right)\ee
where $P_{e_r}= \sqrt{P_{e_r}^I P_{e_r}^I}$ is the gauge invariant momentum. In the limit the $S_{e_r}\rightarrow0$, the  
$P_{e_r}=\frac{2r^2\sin\theta \delta \theta \delta \phi}{r_g^2}$
and $S_{e_r}= 2\delta\theta\delta\phi$. Needless to say in this approximation, we correctly recover $q^{rr}= 1+ O(\tilde t)$. Further, as we consider orbits $r>r_g$, the $|\tilde f(P_{e_r})|$ \cite{adg1} gives
a fractional contribution to the formulas.

Having found the correction to the induced metric in the $\tau=\tau_c$ slice, we then perform a coordinate transformation to find the corrections to the
metric in the Schwarzschild coordinates. The transformations are
\bea
g^{tt} &= &\frac{dt}{d\t}\frac{dt}{d\t} g^{\t\t} + \frac{dt}{dR}\frac{dt}{dR}g^{RR} \ \ \ 
g^{rr}=\frac{dr}{d\t}\frac{dr}{d\t} g^{\t\t} + \frac{dr}{dR}\frac{dr}{dR} g^{RR}\\
g^{rt}&=&\frac{dr}{d\t}\frac{dt}{d\t}g^{\t\t} + \frac{dr}{dR}\frac{dt}{dR} g^{RR}
\eea
We have 
\be
\frac{dt}{d\t}= \frac1{1-f'} \ \  \ \ \frac{dt}{dR}= \pm\frac{f'}{1-f'} \ \ \ \  
\frac{dr}{d R}= \sqrt{\frac{r_g}{r}} \ \ \ \ \frac{dr}{d\tau}= \pm\sqrt{\frac{r_g}{r}} 
\ee
This gives, in particular the corrections to $g^{tr}$ as 
\be
g^{tr}=\pm \frac{2}{1-\frac{r_g}{r}} \left(\frac{r_g}{r}\right)^{3/2} \ \tilde t \ \tilde f\left(\frac{P_{e_r}}{S_{e_r}}\right)
\label{abc}
\ee

The reason we are giving this in details is because the quantum gravity effects have created a $g^{rt}$ term in the corrected metric which normally wouldn't have been
there. The $g_{rt}$ of the inverse metric is $- g^{tr}/(g^{tt}g^{rr})$ and is thus given by the same rhs of (\ref{abc}). This of course diverges at the horizon, but this is a sign of the
failure of the coordinates. The corrections to the cross
terms in the metric $g_{t \phi}, g_{t \theta}, g_{r \phi}, g_{r \theta}, g_{\theta \phi}$ are not there, as by choice of gauge in the internal directions the cross terms like $P^I_{e_r} P^I_{e_{\theta}}=0$ to order $\tilde t$

\section{Semiclassical Effects on Unstable Orbits}
In \cite{cornish}, the unstable orbits in the Schwarzschild space-time appear with Lyapunov exponents. We see how their calculations differ on the inclusion of the quantum gravity corrections. 
Initially, we take linearised corrections $g_{\m \n}+  h_{\m \n}$, and then take the $h_{\m \n}$ as obtained in the previous section. 
 We find the most generalised equation of motion, and the type of unstable orbits which might emerge from the semiclassical perturbations.

 The geodesic equations are derived in their first order form in phase space. The Lagrangian for a particle
is given by (in the squared form) using the conventions which appear in \cite{cornish} (s is a parameter),
\bea
{\cal L} &= &\frac12\left[\left\{-\left(1-\frac{r_g}{r}\right)+ h_{tt}\right\}\left(\frac{dt}{ds}\right)^2 + 2 h_{t r}(r) \frac{dt}{ds} \frac{dr}{d s} + \left\{\frac1{(1-\frac{r_g}{r})}+ h_{rr}\right\}\left(\frac{dr}{d s}\right)^2 \right. \nn \\ 
&+ & \left.\left\{r^2 \sin^2\theta + h_{\phi \phi}\right\}\left(\frac{d\phi}{d s}\right)^2\right]
\eea
The $\theta=\pi/2$ in the above. The Equation of motion are derived from the above as:
\bea
f(r) \left(\frac{dt}{d s}\right) + h_{rt} \left(\frac{dr}{d s}\right) &=& E \label{t}\\
g(r)\left(\frac{d r}{d s}\right) + h_{rt}\left(\frac{dt}{d s}\right)&=& p_r \label{r}
\eea
where $f(r)= - \left(1-\frac{r_g}{r}\right) + h_{tt}$, $g(r)= \left(\frac{r}{r-r_g}\right) + h_{rr}$ and E is constant of motion. In the subsequent equations $f(r),g(r)\equiv f, g$. 

One can solve (\ref{t},\ref{r}) for $dt/d s$ and $dr/d s$ which has to be used in the equation of motion.
\bea
\frac{dt}{d s} &= &\frac{E}{f} - \frac{p_r h_{rt}}{gf} + O(h^2) \\
\frac{dr}{d s}&=& \frac{p_r}{g} - \frac{E h_{rt}}{fg} +O(h^2)
\eea

The equation of motion for $p_r$ is given by $\frac{dp_r}{d s}=\frac{\partial {\cal L}}{\partial r}$ and in addition we make a transformation to the $t$ coordinate
by multiplying $\frac{dp_r}{d s} \frac{d s}{dt}$ and keeping terms to O($h^2$):

\bea
\frac{ d p_r}{d s}\frac{d s}{dt} &= & \frac{1}{2}\left[\partial_r f \left(\frac{dt}{d s}\right)+ 2 \partial_r h_{tr}  \left(\frac{dr}{d s}\right) \right. \\
&+ &\left.\partial_r g(r) \left(\frac{dr}{d s}\right)^2 \left(\frac{d s}{dt}\right)+ \partial_r \tilde q(r) \left(\frac{d\phi}{d s}\right)^2 \left(\frac{ds}{dt}\right)\right]\\
\frac{d p_r}{dt} &=& \left[\frac12 \partial_r f\left(\frac{E}{f} - \frac{p_r h_{rt}}{g f}\right)  
 + \partial_r h_{rt}\left(\frac{p_r}{g}\right) \right. \nn \\
&& \left. + \frac12\partial_r g(r) \left(\frac{p_r^2}{g^2} -\frac{2 E p_r h_{rt}}{fg^2}\right)\left(\frac{E}{f} -\frac{p_r h_{rt}}{gf}\right)^{-1} + \frac12\partial_r \tilde q(r)\frac{L^2}{\tilde q^2} \left(\frac{E}{f} -\frac{p_r h_{rt}}{fg}\right)^{-1}\right]
\eea
 Where $\tilde q\frac{d\phi}{d s}=L$ where $\tilde q = r^2 \sin^2\theta + h_{\phi \phi}$ and L is the angular momentum constant. 
Using the above, one finds the matrix which describes the perturbation flow of the $v\equiv(r,p_r)$  is $\frac{d\delta v_i}{dt}= K_{ij}\delta v_i(t).$
where $i$ runs from 1,2.
Listing the elements of $K_{ij}$, one finds
\bea
K_{rr}&= &\frac{h_{rt}\partial_r \ln f}{g}-\frac{\partial_r h_{rt}}{g}\\
K_{r p_r}&= &\frac{f}{ E g} \\
K_{p_r p_r}&=& - \frac12\partial_r(\ln f)\frac{h_{rt}}{g} + \frac{\partial_r h_{rt}}{g} + \frac12\frac{\partial_r \ln \tilde q L^2 h_{rt} f}{\tilde q g E^2}\\
K_{p_r r}&=& \frac12\partial^2_r\ln f E + \frac12\partial_r\left(\frac{\partial_r\ln \tilde q L^2 f}{ \tilde q E}\right)
\eea
In the above we take the $r=$ const orbits or $p_r= E h_{rt}/f$.
The eigenvalues of the matrix are
\be
\lambda_{1,2} = \frac12[K_{rr} + K_{p_r p_r}] \pm \frac12\sqrt{(K_{rr}+ K_{p_r p_r})^2 -4(K_{rr}K_{p_r p_r} - K_{r p_r}K_{p_r r})}
\label{lya}
\ee
The remarkable fact is that, for orbits which normally have imaginary Lyapunov exponents, there is a real part.
 I thus concentrate on the terms $K_{rr} + K_{p_r p_r}$, which might give detectable effects.
\be
K_{rr}+K_{p_r p_r} =\frac{h_{rt}}2\left[\frac{\partial_r \ln f}{g} + \frac{L^2 f}{E^2 \tilde q g}\partial_r \ln \tilde q\right]
\ee
Using the definition for $g(r)$ and $f(r)$ one finds that
\be
= \frac{h_{rt}}{2}\left[\frac{r_g}{r^2} + \frac{L^2}{E^2}\left(1-\frac{r_g}{r}\right) \frac{2}{r^3}\right]
\ee
This factor is non-zero, positive and real for all orbits $r>r_g$ if we take $g_{rt}=h_{rt}$ for the black hole metric. For the white hole metric,
the Lyapunov exponent will be positive for backwards flow in the time direction. 
{\it Note, the Lyapunov exponent is postive as the quantum fluctuations are negative in the black hole space-time, and the evolution matrix defined as $\dot L_{ij} (t)= K_{im}L_{mj} (t)$  
has a growing perturbation.} The principal Lyapunov exponent is defined in terms of the evolution matrix as: $\lambda= {\rm lim}_{t \rightarrow \infty}  \frac1{2t} \ \ln |L^*_{ji}L_{ij}|.$ 

Thus for a given fixed value of $r, L, E$ we can find $\lambda$. 
If we take a given $L, E$, the radius of the circular orbit in the quantum corrected metric is slightly perturbed. The radius of the circular orbit is calculated
by finding the minimum of the effective potential which appears in the description of the trajectory of an particle. This is obtained in the case of the classical
metric by using the fact that along a time like trajectory, the distance squared is a number ($l$),
\be
\frac{E^2}{1-\frac{r_g}{r}} + \frac{1}{1-\frac{r_g}{r}} \left(\frac{dr}{d s}\right)^2 + \frac{L^2}{r^2} = -l
\label{dist}
\ee
where in the expression for the line element we have put in the constants $E, L$ by replacing the $\frac{dt}{d s}, \frac{d\phi}{d s}$ appropriately.
It follows that this is same as
\be
\left(\frac{dr}{d s}\right)^2 + V(r) =0
\ee
where $V(r)= \left(1-\frac{r_g}{r}\right)\left(\frac{L^2}{r^2} +l\right) + E^2$. The circular orbits are then obtained as a solution for
$\frac{\partial V}{\partial r}=0.$
In case of the corrected metric, the same (\ref{dist}) gets modified to the following:
\be
 \frac{E^2}{f} + g \left(\frac{dr}{d s}\right)^2  + \frac{L^2}{\tilde q} = -l
\ee
The potential is then identified as
\be
V_h(r)= \frac{1}{g} \left[\frac{L^2}{\tilde q} + \frac{E^2}{f} + l \right]
\ee
Solving for $\frac{\partial V_h(r)}{\partial r} =0$ at $r_0+\delta_0$, where $r_0$ extremises the
classical potential

\be
\delta_0= \frac{V(r_0)h'(r_0) - \bar h '(r_0)}{V ''(r_0)}
\label{corr}
\ee
where $' $ denotes the derivative wrt r, and $h(r)= \left(1-\frac{r_g}{r}\right) h_{tt} + \frac{h_{\phi \phi}}{r^2} + \frac{h_{tt}}{1-r_g/r} $
and $\bar h = (1+ L^2/r^2) h_{tt} + \left[\left(1 - \frac{r_g}{r}\right) + E^2\right]h_{\phi \phi}/r^2 $.
However, this correction to the orbit radii does not affect the evaluation of the Lyapunov exponent in the linearized approximation.

Next we plug in the values of $r,L,E$ (\ref{lya}), and find the Lyapunov exponent. We take the same circular orbit considered in \cite{cornish} for the classical Lyapunov exponent in order to see the changes in the quantum corrected metric. 
In units of $GM/c^2$ as in \cite{cornish} we use $L=4, E=1$ and get two orbits at $r_0=4$ and $r_0=12$. In case of $r_0=4$, the classical Lyapunov exponent gets corrected, but to such a small extent, that it is unlikely to manifest itself. Of interest is the $r_0=12$ orbit, which is classically stable.
Is there anyway that the quantum instability will manifest itself? This is a interesting question and we find that for $r_0=12$, the Lyapunov exponent to be: 
(The $l_p= 1.62 \times 10^{-35} m$, $G= 6.67 \times 10^{-11} m^3  kg^{-1} s^{-2}$, $c= 2.99 \times 10^8 m/s$.)
\be
\lambda= \frac{\tilde t c}{10^n l_p} \tilde f(\frac{P_{e_r}}{S_{e_r}})(0.0047)
\ee
The $\tilde t=\frac{l_p^2}{r_g^2}$, and one gets after putting in appropriate dimensional units like the speed of light
\be
\lambda = 10^{40-3n}(8.6)\tilde f\left(\frac{P_{e_r}}{S_{e_r}}\right)/ s
\ee
The radius $r_g$ of the black hole is taken to be $10^{n}l_p$. Thus for astrophysical black holes where $n > 33$, this is a number unlikely to manifest
itself in any experiments. For black holes of horizon radius $ 10^{41/3} l_p\sim 10^{13.67} l_p$ this is order 1. Thus the chaos time
scale for this is $T_{\lambda}\sim 1 s$. But in general, the chaos time scale would be
\be
T_{\lambda}\approx 1.1\times 10^{3n -41} \frac{1}{\tilde f\left(\frac{P_{e_r}}{S_{e_r}}\right)} s
\ee

\section{Detecting the Instabilities}
We investigate the plausible effects of the positive Lyapunov exponent on physical phenomena.
\subsection{Gravity waves and New Instabilities}
The interesting effect of the chaos is in the detection of gravitation waves, as discussed in the papers \cite{cornish}.  
For primordial black holes, the time scale of chaos is found to be $T_{\lambda}= 1/\lambda$ which is order 1s for black holes with horizon radius $10^{13.67} l_p$, and we can make an attempt 
for investigating their effects on gravitational wave forms. The gravitational wave has a time scale
$T_{\omega}= 2\pi/\dot \phi$. Given that in the frame of the observer, $\dot \phi= L(r-2)/E r^3$ for the particular orbit under consideration, we get $T_{\omega}\sim   10^{n-42} s$ which is clearly smaller than the chaos time scale for n=13.67. 
So chaos will not set in. Infact if we take the ratios of the time scales we get
$
\frac{T_{\lambda}}{T_{\omega}}\sim  10^{2n+1 }.$ This is not less than 1 even for n=0 or Planck length black holes which are anyway beyond the semiclassical regime. 
  
We can still try to find the radiative decay rate for a test particle orbiting the black hole, and that is given by
$T_d = \frac{5}{256} \frac{c^5}{G^3}\frac{r^4}{\mu}$,
where $\mu= m_1 m_2 (m_1 + m_2)$ where $m_1$ is the mass of the black hole and $m_2$ is the mass of the test particle. For chaos to have any effect
on the gravitation wave decay rate, we get a bound for the mass of the
test particle: $ m_2 < 8.2 \times 10^{-n-9} kg$. This is obviously a restrictive bound (mass of the electron is $\sim 10^{-31} kg$) , and we are unlikely to find any signatures
of the instability in the gravity wave phenomena for astrophysical black holes (for solar mass black holes $n> 38$). Note that for primordial black holes we should be really
using quantum corrected formulas for `gravity wave emission'. Further, the Lyapunov exponents for null geodesics can be derived as a limiting case of the above calculation,
and these determine the quasinormal decay of a black hole \cite{quasi}. Thus these new semiclassical instabilities will create new quasinormal decay modes for primordial
black holes. 

\subsection{Magnification of quantum gravity corrections to classically unstable orbits.}
As described in the previous section radii of circular orbit 
get modified by the function $\delta_0$ (\ref{corr}).  We find that the corrections to the unstable orbit radii get
magnified in time. To see this, we do the following calculation: 
As per \cite{cornish}, the Lyapunov exponent in appropriate units is $\lambda \sim \frac{1}{1.62(4 \sqrt{2})} 10^{43-n} / s$.
Substituting this in the eom we find that,
using the principal Lyapunov exponent, one has
\be
\delta r(t) = \exp\left( 1.09 \times 10^{42 -n} t\right) \delta_0 
\label{eqn}
\ee

Thus a particle traversing in the quantum gravity corrected metric will start deviating from its trajectory, and this deviation will
grow with time. Let us solve for the time,
\be
t= 9.17\times 10^{n-43} \ln \left(\frac{\delta r(t)}{\delta_0}\right) s 
\ee
Given that $\delta_0 \sim  r_g \tilde t \sim 10^{-n} l_p$ , for $\delta r(t)\sim 1$ cm, we require:
\be
t=9.17 \times 10^{n-43} \ln(10^{n+33}) s = 2.1 \ (n+33) \ 10^{n-42} s 
\ee
Thus, for black holes with horizon radius $10^{42} l_p$ this time is of the order of 1 s. This is in the range of astrophysical black holes,
and is a very significant effect. Thus any test particle traversing in unstable
orbits will start deviating from the classical radii spontanously due to quantum gravity corrections. 
However, in an accreting disk the `instability' due to quantum fluctuations will be difficult to isolate from external matter perturbations.
Thus though the calculations is one of the first attempts to magnify quantum gravity effects to macroscopic length scales, the above
is still a theoretical result, and an appropriate experiment has to be designed to obtain any measurable effect.

\section{Conclusions}
In this article we calculated 
non-zero positive Lyapunov exponents for orbits in a semiclassically corrected metric of a Schwarzschild space-time. However, the Lyapunov
exponents are suppressed by the semiclassical parameter which controls the semiclassical corrections. The instabilities are almost undetectable for astrophysical black holes, and a particle will
have stable motion. For black holes which have horizon radius of $\sim 10^{14} l_p$ the Lyapunov exponent is order $1 s^{-1}$, but
there is no detectable effect of these instabilities for gravitational waves or radiation decay from particles orbiting the
primordial black holes. The instabilities might manifest themselves in other astrophysical phenomena and black hole ring down. We further make the interesting observation that 
due to semiclassical corrections to the metric, the radii of classical circular orbits get corrected. Normally these corrections would be negligible, but the corrections to the classically unstable orbits get magnified in time, and for astrophysical black holes
the deviations grow to order 1 cm in 1 s! This is a spontaneous quantum effect, and not the result of any external perturbations of the black hole metric. However a tangible experiment has to be designed
to make this spontaneous effect detectable in a real accreting situation. Further this effect is based on circular unstable orbits, but the parameter space of orbits has to be
extended, and study of the behaviour of homoclinic orbits which have overlap with the unstable orbits has to be done. 

This calculation can be easily extended to the case of unstable null geodesics, and the Lyapunov exponent of null geodesics appears in the
calculation of the quasi-normal frequency of black hole ring down \cite{quasi}. Moreover, the Lyapunov exponents for Kerr black holes are relevant
for black hole mergers \cite{merge}. The coherent state has to be defined for the Kerr metric in a co-rotating frame, and this is work in progress. 
We expect to extend the calculation of this paper to the case of null geodesics, Kerr black holes and find experiments
which will detect the effects of quantum gravity corrections to the metric.

\noindent
{\bf Acknowledgements:} I would like to thank S. Das and B. Tippet for useful discussions. This research is supported by NSERC and research funds
of University of Lethbridge. I would also like to thank an anonymous referee for pointing to several errors in the previous version of the
manuscript.

\end{document}